\documentclass[aps,showpacs]{revtex4}
\usepackage{amssymb}
\usepackage[dvips]{graphicx}
\usepackage[english]{babel}
\usepackage{indentfirst}
\usepackage{amsxtra}
\usepackage{amsmath}
\usepackage [mathcal]{eucal}
\usepackage{bbold}
\newcommand{\beq}{\begin{equation}}
\newcommand{\eeq}{\end{equation}}
\newcommand{\beqn}{\begin{eqnarray}}
\newcommand{\eeqn}{\end{eqnarray}}
\newcommand{\bsigma}{\mbox{\boldmath $\sigma$}}
\newcommand{\btau}{\mbox{\boldmath $\tau$}}
\newcommand{\half}{\frac{1}{2}}

\newcommand{\br}{{\bf r}}
\newcommand{\bqu}{{\bf q}}
\newcommand{\bpi}{{\bf p}}

\newcommand{\threej}[6]{ \left( \begin{array}{ccc}
                               #1 & #2 & #3 \\
                               #4 & #5 & #6 
                             \end{array}
                        \right) } 
\newcommand{\sixj}[6]{ \left\{ \begin{array}{ccc}
                               #1 & #2 & #3 \\
                                #4 & #5 & #6 
                               \end{array}
                        \right\} } 
\newcommand{\ninej}[9]{ \left\{ \begin{array}{ccc}
                               #1 & #2 & #3 \\
                                #4 & #5 & #6 \\ 
                                #7 & #8 & #9 
                                \end{array}
                        \right\} } 

\usepackage[usenames]{color}
\usepackage{ulem}

\begin{document}

\noindent
\title{Coulomb and spin-orbit interactions in
random phase approximation calculations}

\author{V. De Donno, G. Co'}
\affiliation{Dipartimento di Matematica e Fisica ``E. De Giorgi'',
  Universit\`a del Salento and, 
 INFN Sezione di Lecce, Via Arnesano, I-73100 Lecce, ITALY}
\author{M. Anguiano, A. M. Lallena}
\affiliation{Departamento de F\'\i sica At\'omica, Molecular y
  Nuclear, Universidad de Granada, E-18071 Granada, SPAIN}
\date{\today}

\bigskip
\begin{abstract}
We present a fully self-consistent computational framework composed by
Hartree-Fock plus random phase approximation where the spin-orbit and
Coulomb terms of the interaction are included in both steps of the
calculations. We study the effects
of these terms of the interaction on the random phase approximation
calculations, where they are usually neglected.  We carry out our
investigation of excited states in spherical nuclei of 
oxygen, calcium, nickel, zirconium, tin and
lead isotope chains. We use finite-range effective nucleon-nucleon
interactions of Gogny type.  The size of the effects we find is,
usually, of few hundreds of keV.  There are not simple approximations
which can be used to simulate these effects since they strongly depend
on all the variables related to the excited states, angular momentum,
parity, excitation energy, isoscalar and isovector characters. Even the
Slater approximation developed to account for the Coulomb exchange
terms in Hartree-Fock is not valid in random phase approximation
calculations.
\end{abstract}

\bigskip
\bigskip
\bigskip

\pacs{ }

\maketitle

\section{Introduction}
\label{sec:intro}
The combination of Hartree-Fock (HF) and random phase approximation
(RPA) calculations carried out with a unique
effective interaction has been
able to provide a good description of known nuclear properties in a
wide range of the nuclear chart, from light nuclei around the oxygen
region up to very heavy nuclei such as the uranium. This
success has induced to believe that this computational scheme
could provide good
predictions of the properties of exotic nuclei which will be produced
in the next few years in radioactive ion beams facilities. This
possibility has increased the interest in defining more precisely the
details of the self-consistent HF plus RPA (HF+RPA) calculations.   

In HF calculations, the presence of the spin-orbit term of the
interaction is essential to properly describe the shell structure of
the various nuclei, and that of the Coulomb interaction to distinguish
between proton and neutron single particle (s.p) properties. These two
terms of the effective nucleon-nucleon interaction are usually
neglected in RPA calculations, since the evaluation of their
contributions, considered small as compared to that of the other terms
of the interaction, is computationally quite heavy.

The relatively small size of the effects of Coulomb and spin-orbit terms 
has been confirmed in recent years by the results of some fully self-consistent
HF+RPA calculations. The calculations carried out
with zero-range Skyrme forces \cite{ter05,fra05,sil06,col07,los10,col13}  
indicate that spin-orbit and Coulomb interactions
produce effects of a few hundreds of keV. 

To the best of our knowledge, fully self-consistent HF+RPA
calculations with finite-range interactions have been carried out only
by using Gogny interactions. In this type of calculations, the results
obtained by P\'eru et al. \cite{per05} 
show that the spin-orbit term of the
interaction plays a remarkable role in the structure of the
low-lying quadrupole and octupole states by modifying both excitation
energies and transition probabilities. In the same work it has been shown
that the Coulomb force in RPA calculations significantly affects the
centroid energies of the isovector giant dipole resonances and 
their energy weighted sum rule values.  
The study of Ref. \cite{per05} has ben conducted by considering the
doubly magic nuclei $^{78}$Ni, $^{100}$Sn, $^{132}$Sn, and 
$^{208}$Pb with the D1S$^\prime$ parameterization of the Gogny two-body
effective interaction.

Recently, we have
developed an approach to carry out HF+RPA self-consistent
calculations with finite-range interactions 
\cite{don11a}.  
We have used this model  
to study magnetic and electric nuclear excitations
with Gogny interactions, but in these
investigations the spin-orbit and Coulomb terms of the interactions were
not considered in the RPA calculations.  

In the present work, we show
the results of a study in which we have evaluated the effects of 
these terms of the D1M parameterization of the Gogny interaction in fully
self-consistent HF+RPA calculations. With respect to the investigation
of Ref. \cite{per05}, we have considered a 
different interaction, a wider set of spherical
nuclei, and we have focussed our attention mainly to low-lying
excited states, 
rather than to the centroid energies of giant resonance
excitations. We have studied the validity of the Slater approximation
\cite{sla51} in the treatment of the Coulomb exchange RPA terms. We
have calculated the effects of the spin-orbit and Coulomb terms 
on low lying $2^+$ and $3^-$ multipole excitations and the dependence of these
effects in isoscalar (IS) and isovector (IV) excitations in nuclei
with the same number of protons and neutrons. We have considered
excitations dominated by single
particle-hole (p-h) pairs and we
have studied the evolution of the effects with different values of the
angular momentum of the excitations. Our results confirm that the
effects of the spin-orbit and Coulomb terms of the interactions are of
a few hundreds of keV.

Our model is presented in Sect. \ref{sec:model}.  Details and basic
ingredients of the calculations are presented in
Sect. \ref{sec:details}. In Sect. \ref{sec:results} we have discussed 
the effects of the spin-orbit and Coulomb interactions in a selected
set of results and studied the validity of the Slater
approximation for the Coulomb exchange term. In
Sect. \ref{sec:conclusions} we summarize the main results of our work
and we draw our conclusions.

\section{The model}
\label{sec:model}

The only input required by our self-consistent approach
is the effective nucleon-nucleon force. 
We have considered a general finite-range force which we 
express as
\beq
\widehat{V}_\alpha (i,j)\,= \, v_\alpha(r_{ij}) \, O^\alpha_{i,j} 
\, , \,\,\, \alpha=1,2,\ldots,8 
\, ,
\label{eq:force1} 
\eeq
where $v_\alpha$ are scalar functions of the distance between the two
interacting nucleons, and $O^\alpha$ indicates the type of operator
dependence
\beqn
O^{\alpha}_{i,j}&:& 1\,,\,\,\btau(i)\cdot\btau(j)\,,\,\,
                 \bsigma(i)\cdot\bsigma(j)\,,\,\,
 \bsigma(i)\cdot\bsigma(j)\,\btau(i)\cdot\btau(j)\,, \nonumber
\\ &~&
S_{ij}\,,\,\, S_{ij}\, \btau(i)\cdot\btau(j)
\,,\,\, {\bf L}_{ij} \cdot {\bf S}  
\,,\,\, {\bf L}_{ij} \cdot {\bf S}\, \btau(i)\cdot\btau(j)  
\label{eq:fcahnnels}
\, .
\eeqn 
In the above expression 
$\bsigma$ is the Pauli matrix operator acting on the spin
variable and 
$\btau$ the analogous operator for the isospin.
The tensor operator is defined as
\beq
S_{ij}\, = \,3 \, \frac {\bsigma(i)\cdot \br_{ij} \,
                    \bsigma(j)\cdot \br_{ij} } 
                 {r_{ij}^2}\,
- \, \bsigma(i)\cdot\bsigma(j) \, ,
\eeq
where
\beq
\br_{ij}Ê\, = \, \br_i \, - \, \br_j 
\label{eq:r}
\eeq
represents the relative coordinate.
In the spin-orbit terms of the force, $\alpha=7,8$,
we have indicated with
\beq
{\bf L}_{ij} \,= \, \br_{ij} \times \bpi_{ij}
\label{eq:L} 
\eeq
the relative
angular momentum of the two interacting nucleons, 
where their relative momentum has been
defined as 
\beq
\bpi_{ij}\, = \, \frac{1}{2}(\bpi_i -\bpi_j)
\, ,
\label{eq:p}
\eeq
and with 
\beq
{\bf S}\, = \, {\bf s}_i + {\bf s}_j
\label{eq:S}
\eeq
the total spin of the nucleon pair.

With this type of interactions, we solved the HF equations as
indicated in Refs. \cite{co98,bau99}. From the solution of these
equations we obtained a set of s.p. wave functions that have been used
to solve the RPA equations.  We have considered the RPA equations in
their matrix formulation \cite{fet71,rin80,suh07}.

We have evaluated the corresponding matrix elements by expressing the force in
configuration space as the Fourier transform of the force given in
momentum space
\beq
v_\alpha(r_{ij})=\frac{1}{(2\pi)^{3/2}}\int {\rm d}^3q \, 
\exp \left[i\bqu \cdot (\br_i-\br_j)\right]\, \tilde{v}_\alpha(q)\, .
\label{force:ft}
\eeq
In this way, we could separate the coordinates $\br_i$ and $\br_j$ and 
carry out the multipole expansion of the two exponentials. 
 A detailed derivation of the matrix element expressions for all the
force channels up to $\alpha=6$ can be found in Ref. \cite{don08t}. 

In our previous works \cite{don11a}, the
spin-orbit and Coulomb matrix elements have been neglected in RPA
calculations, and we have considered them in HF calculations only.
The inclusion of the Coulomb interaction in the RPA calculations
\beq
v_{\rm C}(r_{ij})\,=\,\frac{e^2}{|\boldsymbol r_i-\boldsymbol r_j|}
\label{eq:coulomb}
\eeq
is relatively easy, since the RPA matrix elements are identical to
those of the scalar term of the interaction (\ref{eq:force1}) (see
Ref. \cite{don08t}).  Obviously, we have to consider that the
interaction is active only between proton p-h pairs.

The evaluation of the spin-orbit matrix elements is
more involved. We give in Appendix
  \ref{sec:appa} some details about it.  The general expressions
presented in this appendix have been obtained by considering that the scalar functions $v_{\alpha=7,8} $ of
Eq. (\ref{eq:force1}) have finite-range. In our calculations
we used Gogny interactions that include
a spin-orbit potential of contact type, analogous to that adopted in Skyrme-like interactions:
\beq
F_{ij}^{\rm SO} \, =\, 2\, i\, W_0 \,
\left[ \overleftarrow{\bpi}_{ij} \times \delta(r_{ij}) 	\, \overrightarrow{\bpi}_{ij} \right]
\cdot \mathbf{S} \, ,
\label{eq:vLSgogny}
\eeq
where the arrows indicate the side on which the operator $\bpi_{ij}$ 
acts. Taking into account the expression 
\beq
\delta(r_{ij})\, =\, \lim_{\mu \to \infty} \, 
      \frac{\mu^3}{\pi^{3/2}} \, \exp \left(- \mu^2 r_{ij}^2 \right) \, ,
\eeq
we obtain:
\beqn
F_{ij}^{\rm SO}&=& 2\, i\, W_0 \, \lim_{\mu \to \infty}
\frac{\mu^3}{\pi^{3/2}} \left[ \overleftarrow{\bpi}_{ij}  \times \exp \left(- \mu^2 r_{ij}^2 \right)
   \, \overrightarrow{\bpi}_{ij} \right] \cdot \mathbf{S} \nonumber\\
&=&-\,4\,W_0 \, \left[ \lim_{\mu \to \infty} \, \frac{\mu^5}{\pi^{3/2}} \, \exp \left(- \mu^2 r_{ij}^2 \right)
\right] \, \mathbf{L}_{ij} \cdot \mathbf{S}
\, .
\eeqn
By comparing the above expression with the 
$v_7$ term of
Eq. (\ref{eq:force1}), we identify
\beq
v_7(r_{ij})\, =\,-\,4\,W_0 \, \lim_{\mu \to \infty} \, \frac{\mu^5}{\pi^{3/2}} \, \exp \left(- \mu^2 r_{ij}^2 \right)
\, 
\eeq
whose Fourier transform is 
\beq
 \tilde{v}_7(q)\,=\, \lim_{\mu \to \infty}\,
\int {\rm d}^3r \, 
\exp \left[-i\bqu \cdot (\br_i-\br_j)\right]\, v_7(r)\,
= \,W_0 \, \bqu^2 \, ,
\label{force:ft7}
\eeq
which is the expression used in our RPA calculations.

In the following, we indicate with $\omega_0$ the excitation energies
obtained without spin-orbit and Coulomb interactions in RPA calculations. In analogy,
we call $\omega_{\rm C}$, $\omega_{\rm SO}$, and  $\omega_{\rm C+SO}$ the energies obtained
when, only the Coulomb, or only the spin-orbit term, or both are 
included. 

\section{Details of the calculations}
\label{sec:details}

The results we present in this article have been obtained by using the
D1M parameterization \cite{gor09} of the Gogny interaction
\cite{dec80}. We carried out calculations also with the more
traditional D1S force \cite{ber91} but, since the results are very
similar to those obtained with the D1M interaction, we do not show and
discuss them here. The D1M interaction is composed by four
finite-range terms, the scalar, isospin, spin and spin-isospin
dependent terms, a zero-range density dependent term and, in addition,
the Coulomb and a zero-range spin-orbit term.

The first step of our calculations consists in constructing the
s.p. basis by solving the HF equations 
with the complete D1M interaction described above.
This is done by imposing bound-state boundary conditions at the edge
of the discretization box.  The technical details concerning the
iterative procedure used to solve the HF equations for a
density-dependent finite-range interaction can be found in
Refs. \cite{co98b,bau99}.  When the stable solution, corresponding to
the minimum of the binding energy, is reached, the HF equations are
solved again, not only for the states below the Fermi surface, but, by
using the local Hartree and the non local Fock-Dirac potentials
constructed on these s.p. states, also for those states above it. 
In this manner we generate a set of discrete
bound states also in the positive energy region, which should be
characterized by the continuum. The level density in the continuum
region is strictly related to the size of the space integration box:
the larger is the box the higher is the level density.

We write the RPA secular equations \cite{rin80} in matrix form and
solve them by diagonalization. The dimensions of the matrix to
diagonalize are given by the number of the p-h pairs contributing to
the specific excitation. This depends on the number of the s.p. states
composing the configuration space. In our approach, the results of the
RPA calculations depend on two parameters, the level density, which is
related to the size of the integration box, and the maximum
s.p. energy.  We have chosen the values of these two parameters by
controlling that the centroid energies of the giant dipole responses
do not change by more than 0.5 MeV when either the box size or the
maximum s.p. energies are increased.  The most demanding calculations
are those we carried out for the $^{208}$Pb nucleus.  In this nucleus,
by using a box radius of 25 fm and an upper limit of s.p. energy of
100 MeV, we diagonalise matrices of dimensions of about 1300 $\times$
1300.

%%%%%%%%%%%%%%  
\section{Results}
\label{sec:results}
In this section we study the effects of the Coulomb and spin-orbit
interactions in RPA calculations. 
For this study we have considered a set of isotopes representative of various
regions of the nuclear chart. For the light nuclei we have chosen the 
oxygen isotopes $^{16}$O,  $^{22}$O,  $^{24}$O,  $^{28}$O, 
for the medium nuclei some calcium, 
$^{40}$Ca, $^{48}$Ca, $^{52}$Ca, $^{60}$Ca,  
and nickel isotopes, $^{48}$Ni, $^{56}$Ni, $^{68}$Ni, $^{78}$Ni, 
for the heavier nuclei some tin isotopes, 
$^{100}$Sn, $^{114}$Sn, $^{116}$Sn, $^{132}$Sn, 
and, in addition, the
$^{90}$Zr and $^{208}$Pb nuclei. 
Common feature of all
these nuclei is that the s.p. levels below the Fermi surface are
fully occupied, and those above it are completely empty. This implies
that the nuclei we have considered have spherical shape. In addition,
since the energy gap between the last occupied s.p. level and the
first empty level is relatively large, the pairing effects are
negligible. Our calculations do not consider these effects, 
even though the 
Hartree-Fock-Bogoliubov calculations of Ref. \cite{hil07} indicate 
the presence of pairing effects in the 
$^{22}$O, $^{52}$Ca, $^{60}$Ca, $^{68}$Ni, $^{90}$Zr, $^{114}$Sn 
and $^{116}$Sn nuclei. 

 First, we have focussed our
attention into low-lying quadrupole and octupole electric excitations,
more precisely those $2^+$ and $3^-$ excitations dominated by p-h
pairs where the particle is below the continuum threshold. 
We found  $3^-$ states with these characteristics for all the nuclei we
have investigated. On the contrary, these type of states are not present in
the $2^+$ excitations of $^{16}$O, $^{28}$O, $^{40}$Ca, $^{60}$Ca and
$^{48}$Ni.
We have
studied the effects of the Coulomb force, and the need of an exact
treatment of its exchange matrix elements (Sect. \ref{subsec:slater}),
and then the effects of the spin-orbit interaction
(Sect. \ref{subsec:2+3-}). In section \ref{subsec:ISIV} we have
investigated whether the Coulomb and spin-orbit interactions have
different effects on IS and IV excitations in nuclei with $N=Z$. In
section \ref{subsec:pstate} we have studied the sensitivity to Coulomb
and spin-orbit terms of those states dominated by a unique
s.p. transition, as a function of the angular momentum of the
excitation.

\subsection{Effects of the Coulomb force on quadrupole and octupole
  electric excitations} 
\label{subsec:slater}

The use of a finite-range interaction in a fermionic many-body system
requires the evaluation of both direct and exchange matrix elements of
the force.  The evaluation of these latter ones
is numerically much more involved than that of the former ones. 
For this reason, the exchange matrix
elements of the Coulomb interaction are often estimated by using a
local density approximation, called Slater approximation \cite{sla51},
which reduces their contribution to a correction of the direct matrix
elements.
The validity of the Slater approximation in
HF calculations has been investigated for the zero-range Skyrme
force \cite{tit74, ska01, blo11} but also for the finite-range Gogny
\cite{ang01a} interaction. 

We study the validity of the Slater approximation in RPA
   calculations by comparing its results with those obtained
   by exactly evaluating the Coulomb exchange matrix elements.
The expression of the Coulomb exchange contribution to the total HF
energy in the Slater approximation is \cite{col13}
\beq
E_{\rm C,ex}^{\rm Slater} \,=\, -\frac{3e^2}{4}\, \left( \frac{3}{\pi} \right)^{ 1 / 3 }\,
\int {\rm d}^3 r \, \rho_p^{ 4 / 3} ( r ) \, ,
\eeq
where $\rho_p$ is the proton density. From the above equation we
obtain 
\beq
V_{\rm C,ex}^{\rm Slater}(r_{ij})\, =\, -\frac{e^2}{3} \, \left( \frac{3}{\pi} \right)^{1 / 3}
\rho_p^{- 2 / 3} ( r_j ) \, \delta(r_{ij}) \, ,
\eeq
which is the exchange Coulomb potential in the Slater approximation
to be used in RPA calculations \cite{col13}. 
Specifically, we added this
expression to that of the Coulomb potential (\ref{eq:coulomb}), and
calculated only the direct matrix elements.

To study the effects of the Coulomb interaction on the RPA, we use the
same set of s.p. states generated by the HF calculations with the full
D1M Gogny interaction, and we compare the results obtained without
Coulomb interaction with those obtained by including it in the exact
manner and in the Slater approximation. In this study the spin-orbit term
of the interaction has not been considered in the RPA calculations.

\begin{figure}
\begin{center}
\includegraphics[width=8cm] {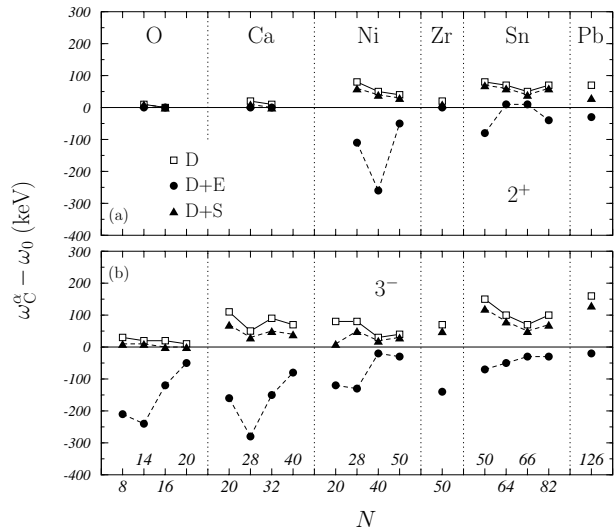}
%\vspace{-2.cm}
\caption{\small Difference between the RPA energies calculated
  with and without Coulomb interaction for quadrupole and octupole
  electric excitations of the nuclei we are investigating. 
  The open squares, $\alpha \equiv$D indicate the results obtained by
  considering the Coulomb direct term only, the solid circles, 
  $\alpha \equiv$D+E those obtained by considering also the exchange term.
  The solid triangles, $\alpha \equiv$D+S, show the results
  obtained by using the Slater approximation to describe the exchange
  term. All the results of the figure have been obtained
  without spin-orbit interaction in
  the RPA calculation. The lines  are drawn to guide the eyes. 
  }
\label{fig:coulomb}
\end{center}
\end{figure}

For the various nuclei we have investigated, 
we present in Fig. \ref{fig:coulomb} the RPA results for low-lying 
$2^+$ and $3^-$ states, panels (a) and (b) respectively, 
carried out by including the Coulomb interaction. The results presented are the
differences between the energies obtained with and without
Coulomb interaction, $\omega_{\rm C}^\alpha-\omega_{0}$.  
The open squares, $\alpha\equiv  {\rm D}$, indicate the results
obtained when only the direct terms of the Coulomb matrix elements are
considered, and the solid circles, $\alpha\equiv {\rm D+E}$, when also
the exchange matrix elements are included. The solid triangles,
$\alpha\equiv {\rm D+S}$, show the results obtained when the Slater
approximation of the exchange matrix elements is used. 

The results shown in Fig. \ref{fig:coulomb} indicate that the effects
of the Coulomb interaction are rather small. We observe maximum
differences of the order of few hundreds of keV, in much cases smaller
than 100 keV. If only the direct matrix elements are considered the
Coulomb interaction is always repulsive: all the nuclei show positive
differences, smaller than 100 keV for the $2^+$, and smaller than 200
keV for the $3^-$. The sign of the difference is reversed when the
exchange terms are considered, as the solid circles indicate. The
behavior of the complete results strongly depend on the multipolarity
and on the nucleus considered. The effects on the $2^+$ states of the
oxygen and calcium isotopes are negligible, while they become
remarkable in the nickel isotopes, more relevant than the effects
found in the heavier nuclei we have considered. The situation on the
$3^-$ states is again different. In this case, the isotopes where we
observe the largest effects are those of oxygen and calcium, while the
effects on the heavier nuclei become gradually smaller.

The results obtained with the Slater approximation strictly follow
those obtained by considering only the direct term, and slightly lower
the size of the repulsive effect. The Slater approximation is unable
to modify the effects of the direct Coulomb matrix elements to
reproduce in a reasonable way the effects of the exchange terms.

%%%%%%%%%%%%%%%%%%%%%%%%%%%%%
\subsection{Effects of the spin-orbit force on quadrupole and octupole
  electric excitations} 
\label{subsec:2+3-}

\begin{figure}
\begin{center}
\includegraphics[width=8cm] {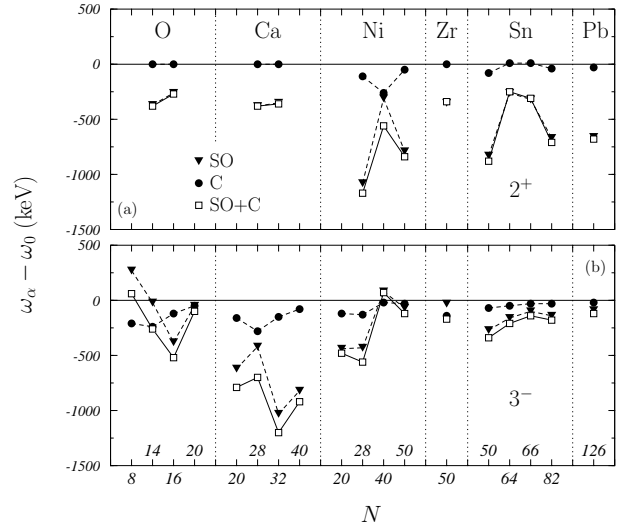} 
%\vspace{-2.cm}
\caption{\small The same as in the previous figure, but now the 
 various symbols represent the RPA energy differences obtained 
 by considering the spin-orbit interaction only (solid
  triangles), the Coulomb interaction only
 (solid circles) and both of them (open squares). 
 As in Fig. \ref{fig:coulomb} these results are expressed as 
 differences with the RPA energies calculated 
 without these terms of the interaction. 
  The lines  are drawn to guide the eyes. 
}
\label{fig:2+3-}
\end{center}
\end{figure}

We use again the $2^+$ and $3^-$ states considered in the previous
section to discuss the effects of the spin-orbit interaction.  The
results of the calculations done by including the Coulomb and
spin-orbit terms of the interactions for the evaluation of the
excitation energy of these multipoles are presented in
Fig. \ref{fig:2+3-} as a difference with the energies $\omega_0$ 
obtained without them.

In Fig. \ref{fig:2+3-} the solid triangles indicate the results obtained
by using the spin-orbit force only, 
while the results of the complete calculations, 
where both Coulomb and spin-orbit terms are considered, are shown by 
the open squares. For completeness, we show again, with the solid circles,
the results obtained by using the Coulomb interaction only. 

We first remark that the effects of the spin-orbit force are, in the
great majority of the cases, larger than those of the Coulomb
interaction.  The second remark is that, in general, the spin-orbit
interaction is attractive.  The exceptions to this trend that we
observe are for the $3^-$ excitations of $^{16}$O and of $^{68}$Ni.
The global effect is essentially given by the simple sum of the two
effects separately considered. The largest effects are those found for
the $3^-$ state in $^{52}$Ca nucleus and for the $2^+$ state in
$^{56}$Ni nucleus where they reach the values of about 
1.2 MeV. A comparison of our results with those of P\'eru et
al. \cite{per05} for the $^{78}$Ni, $^{100}$Sn, $^{132}$Sn and
$^{208}$Pb show a good agreement.

\begin{figure}[b]
\begin{center}
\includegraphics[width=8cm] {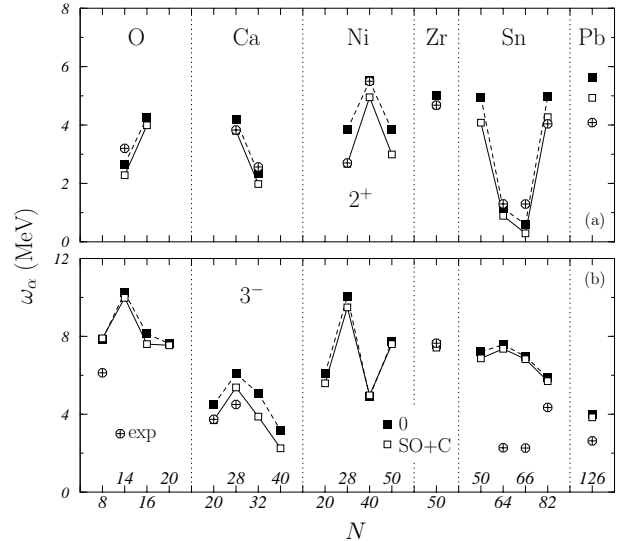}
\caption{\small Excitation energies of the lowest collective $2^+$ (a)
  and $3^-$ (b) states for the nuclei we are investigating. 
  The solid squares indicate the results 
  obtained by neglecting both Coulomb and spin-orbit interactions 
  and the open squares those obtaind by considering them. 
  These energies are compared with the available experimental 
  values taken from Refs. \cite{led78,bnlw} and  
  represented by the crossed circles. The lines  are drawn to guide
  the eyes. 
}
\label{fig:energy0}
\end{center}
\end{figure}

In Fig. \ref{fig:energy0} we show the energies of the $2^+$ and $3^-$
states obtained by using the D1M interaction with and without Coulomb
and spin-orbit terms, open and solid squares respectively.  We
  observe that the inclusion of the Coulomb and spin-orbit terms
  reduces the energy values. We compare the
results of our calculations with the available experimental
values taken from the compilations of Refs. \cite{led78,bnlw,ram01,sto05} 
(crossed
circles). The experimental spectrum is much richer than that
  produced by our calculations, therefore, in some case, the
identification of the experimental excited state to be compared with
that theoretically found is not free from ambiguities. For this reason, 
we do not enter in a detailed discussion of each result.
In any case, we can state that the comparison shown in
Fig. \ref{fig:energy0} is satisfactory, especially considering that
these RPA calculations are parameter free.  

The case of the low-lying 2$^{+}$ state in $^{208}$Pb has been 
carefully investigated. 
Calculations carried out with SLy4 interaction \cite{col07} 
and with the Gogny D1S$^\prime$ force \cite{per05} indicate that the
inclusion of spin-orbit and Coulomb terms reduces the discrepancy 
between theory and experiment.
We observe the same effect also in our calculations, where 
energy value of 5.61 MeV is reduced to 4.93 MeV when 
both terms are included. The experimental value is 4.08 MeV
(Ref. \cite{zie68}) and the shift of about 0.85 MeV we obtain in 
our calculation is similar to that found in Refs. \cite{per05,col07}.

The differences observed for the $3^-$ states with respect to the
experimental values are smaller than 2 MeV except for $^{114}$Sn and
$^{116}$Sn. In these nuclei, the inclusion of pairing effects improves
the agreement \cite{ans06,car12}.

\subsection{IS and IV excited states}
\label{subsec:ISIV}

In this section we investigate whether the Coulomb and spin-orbit
forces generate different effects on IS and IV excitations. For this
study, we have selected cases where the IS and IV characters of the
multipole excitation are well identified.  We have limited our
investigation to nuclei with equal number of protons and neutrons and
to multipole excitations dominated by the p-h pairs where particle and
hole states are just above and just below the Fermi surface.  
In this situation, we can identify excited states dominated by
  the same p-h pairs in both proton and neutron sectors. 
  The quantum numbers identifying the s.p. states of these p-h pairs 
  are the same for protons and neutrons.
When the proton
and neutron p-h pairs are in phase we have an IS excitation and when
they are out of phase we have the IV excitation.  Thus, the IS and IV
character of the excitation can be easily identified in our RPA
calculations by observing the relative sign of the proton and neutron
RPA forward $X$ amplitudes 
defined, as usual \cite{rin80}, as
\beq
| \nu \rangle\, = \, \sum_{ph} \left( X^\nu_{ph}\, a^+_p\, a_h \,- \, Y^\nu_{ph}\,  a^+_h \, a_p \right)  |0 \rangle \, ,
\eeq
where $|\nu \rangle$ and $|0 \rangle$ are the RPA excited and ground
states, and $a^+$ and $a$ the creation and annihilation
s.p. operators.

\begin{table}[htb]
\begin{center}
\begin{tabular}{cccccccc}
\hline \hline
&&& \multicolumn{5}{c}{excitation energy (MeV)} \\\cline{4-8}
&&& \multicolumn{2}{c}{IS} &~~~& \multicolumn{2}{c}{IV} \\ \cline{4-5}\cline{7-8}
   nucleus &   p-h pair &  $J^\pi$ & $\omega_0$ & exp  && $\omega_0$ & exp \\
\hline
 $^{16}$O     &   $1d_{5/2}\,1p_{1/2}^{-1}$ & $2^-$ & 9.42 & 8.87 && 11.25 & 12.53\\
      &   $1d_{5/2}\,1p_{3/2}^{-1}$ & $4^-$ & 16.0 & 17.79 &&16.98 & 18.98 \\\hline
 $^{40}$Ca   &   $1f_{7/2}\, 1d_{3/2}^{-1}$  & $2^-$ & 6.51& 7.53 && 8.38 & 8.42 \\
    &   $1f_{7/2}\,1d_{3/2}^{-1}$  & $4^-$ & 6.66 & 5.61 && 7.03 & 7.66 \\\hline
 $^{56}$Ni    &   $1f_{5/2}\, 1f_{7/2}^{-1} $ & $1^+$ & 7.93 &&  & 11.16 & \\
     &  $2p_{3/2}\, 1f_{7/2}^{-1} $ & $3^+$ & 5.79 && & 6.21 & \\
      &  $2p_{3/2}\, 1f_{7/2}^{-1} $ & $5^+$ & 5.99 && & 6.30  & \\\hline
 $^{100}$Sn    &  $1g_{7/2}\, 1g_{9/2}^{-1} $ & $1^+$ & 7.62 && &10.37 & \\
     &  $2d_{5/2}\, 1g_{9/2}^{-1} $ & $3^+$ & 6.53 && & 7.00  & \\
     &  $2d_{5/2}\, 1g_{9/2}^{-1} $ & $5^+$ & 6.59 && & 6.88  & \\
\hline\hline
 \end{tabular}
\caption{\small Isotopes with equal number of protons and neutrons selected
  to study the effects of Coulomb and spin-orbit interactions on IS
  and IV excitations. In the second column we indicate 
  the proton and neutron p-h pairs which dominate the excitations, 
  and in the third column the angular momentum and the parity of the
  excited states we have considered. The values, in MeV, of the excitation
  energies $\omega_0$, are those obtained without Coulomb and spin-orbit 
  interactions. When available, we present in the 
  columns labelled ``exp'' the experimental values taken from the
  compilations of Refs. \cite{led78,bnlw}.
}
\label{tab:ISIV}
\end{center}
\end{table}

\begin{figure}
\begin{center}
\includegraphics[width=8cm] {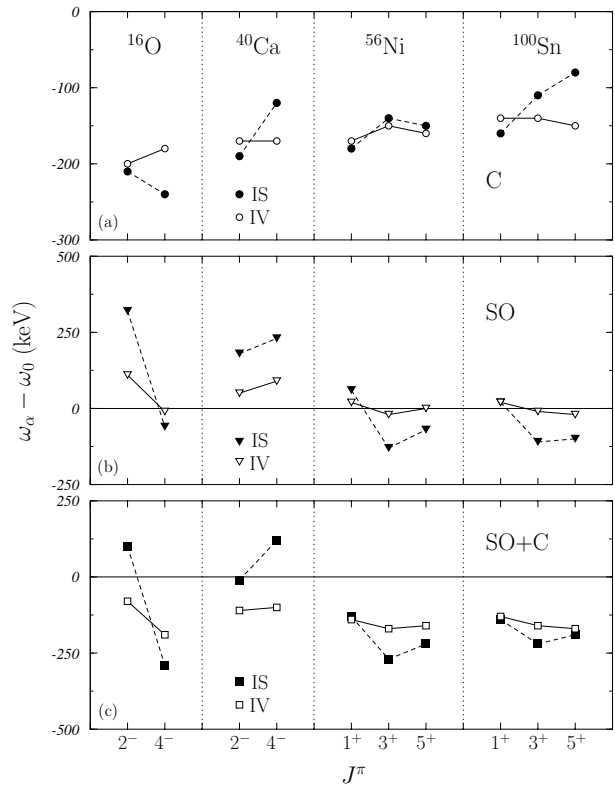}
%\vskip -4.0 cm 
\caption{\small Differences between excitation energies obtained with, 
$\omega_\alpha$, and without, $\omega_0$, Coulomb and spin-orbit
  interactions. In the panel (a) $\omega_\alpha$ has been obtained by
  including the Coulomb interaction only, in the panel (b) the spin-orbit interaction only, and in the panel (c) both of them. 
  The isotopes and the excited states are those listed in Table 
  \ref{tab:ISIV}. In each panel, 
  the solid and open symbols indicate the IS and IV results,
  respectively. The lines are drawn to guide the eyes. 
}
\label{fig:ISIV}
\end{center}
\end{figure}

In Table \ref{tab:ISIV} we show the nuclei we have considered, the p-h
pairs dominating the transitions, and the angular momentum and parity
of the multipole excitation. We also indicate the values of the
energies $\omega_0$ obtained without the Coulomb and spin-orbit
interactions. We compare these energies with the available 
experimental energies taken from the compilations of
Refs. \cite{led78,bnlw,ram01,sto05}. The information given in Table
  \ref{tab:ISIV} completes that given in 
Fig. \ref{fig:ISIV}, where
the effects of the Coulomb and spin-orbit interaction terms are shown 
as differences with respect to $\omega_0$. 
The labels C, SO and SO+C indicate the results we have obtained 
by adding to the D1M interaction 
only the Coulomb, only the spin-orbit or both terms of the 
interactions, respectively.

The results shown in the panel (a) of 
Fig. \ref{fig:ISIV} indicate that the Coulomb interaction is always
lowering the excitation energy values,
and its effects are rather insensitive to the IS or
IV nature of the excitation. 
In the panel (b) of Fig. \ref{fig:ISIV}, we observe ,
that the effects of the spin-orbit 
interaction are, in absolute value, always larger in the
IS than in the IV excitations. 
The larger difference between the effects of the spin-orbit interaction
on IS and IV excitations
occurs for the $2^-$ state in $^{16}$O. 
The sign of the spin-orbit effects is not always the same. 
We observe an enhancement of the energy values for all the 
$1^+$ and $2^-$ states shown in the figure, 
and also for the $4^-$ state in $^{40}$Ca.
For the other cases the energy values are lowered.

We show in the panel (c) of Fig. \ref{fig:ISIV} the global effect
obtained by considering both Coulomb and spin-orbit interactions.  In
the case of IV excitations, we observe a reduction of the energy
values of about 150 keV, almost independent of the
multipolarity and nucleus considered.  The situation for the IS states
is more complicated. We find a lowering of the $\omega_0$ values 
in all cases except for the $2^-$ state
in $^{16}$O and for the two states we have considered in $^{40}$Ca.
In any case, the size of these effects is rather small, reaching
$\sim 300$ keV at most. The inclusion of these terms slightly worsen
the agreement with the available experimental data, as it is
possible to deduce from the results shown in Table \ref{tab:ISIV}. 

%%%%%%%%%%%%%%%%%%%%%%%%%%%%%
\subsection{Excited states dominated by a specific s.p. transition}
\label{subsec:pstate}

In this section we investigate how the effects of the  
Coulomb and spin-orbit interactions depend on
the angular momentum of the excitation.
For this study  
we have selected p-h transitions involving
s.p. states near the Fermi energy. When it has been possible
we have chosen cases where the particle state is 
below the continuum threshold. We have calculated all the multipole
excitations compatible with the angular momentum coupling
of the dominant p-h pair. 

\begin{figure}
\begin{center}
\includegraphics[width=8cm] {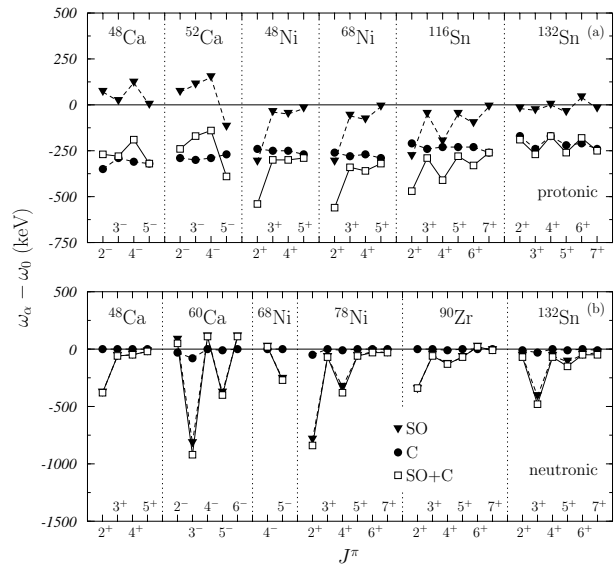}
\caption{\small Energy differences between results obtained 
 by selectively including the Coulomb and spin-orbit interactions, 
 and those without them. 
 The various lines connect the results obtained by 
 including only the Coulomb interaction (solid circles), only the 
 spin-orbit interaction (solid triangles) and both of them (open
 squares). The p-h pairs dominating the various excitations are 
 indicated in Table \ref{tab:sp}.  The lines are drawn to guide the
 eyes.  
}
\label{fig:pnmult}
\end{center}
\end{figure}

\begin{table}[htb]
\begin{center}
\begin{tabular}{ccccc}
\hline \hline
&~~~& \multicolumn{3}{c}{p-h pair} \\ \cline{2-5}
                  &      &  protons  &~~ & neutrons \\
\hline
 $^{48}$Ca     & &   $1f_{7/2}\, 1d_{3/2}^{-1}$ & &  $2p_{3/2}\, 1f_{7/2}^{-1}$ \\
 $^{52}$Ca     & &   $1f_{7/2}\, 1d_{3/2}^{-1}$ &&   \\
 $^{60}$Ca     & &    & &  $1g_{9/2}\, 1f_{5/2}^{-1}$ \\ \hline
 $^{48}$Ni      & &   $2p_{3/2}\, 1f_{7/2}^{-1}$ & & \\
 $^{68}$Ni      & &   $2p_{3/2}\, 1f_{7/2}^{-1}$ & &  $1g_{9/2}\, 2p_{1/2}^{-1}$ \\
 $^{78}$Ni      & &    & &  $2d_{5/2}\, 1g_{9/2}^{-1}$ \\ \hline
 $^{90}$Zr      & &    & &  $2d_{5/2}\, 1g_{9/2}^{-1}$ \\ \hline
 $^{116}$Sn   & &   $2d_{5/2}\, 1g_{9/2}^{-1}$ & & \\
  $^{132}$Sn  & &  $2d_{5/2}\, 1g_{9/2}^{-1}$ & &  $2f_{7/2}\, 1h_{11/2}^{-1}$ \\
\hline \hline
\end{tabular}
\caption{\small
Particle-hole pairs dominating the excited states
considered in Fig. \ref{fig:pnmult}. 
}
\label{tab:sp}
\end{center}
\end{table}

A selection of the most relevant results of this investigation
is shown in Fig. \ref{fig:pnmult} as difference
between the energies obtained by including Coulomb and  
spin-orbit terms with those obtained without them.  We present 
in Table \ref{tab:sp} the p-h pairs dominating the excitations 
of each nucleus considered in the figure. 

We show
in the panel (a) of Fig. \ref{fig:pnmult} the results regarding the 
multipole excitations dominated by proton s.p. pairs. The solid
circles indicate the results obtained by considering
only the Coulomb interaction, the solid triangles those obtained by
considering only the spin-orbit interaction and the open squares those
where both interactions have been considered. 

We observe that effects of the Coulomb interaction are 
essentially independent of the multipole excitation, and produce 
always a lowering of the energy values.

The situation is much more complex for the results obtained with the
spin-orbit interaction. For example, in $^{48}$Ca and $^{52}$Ca
isotopes we observe an enhancement of the $2^-$, $3^-$ and $4^-$
excitation energies, and a lowering of the energies of the other
multipoles considered. We do not identify general trends related to
the change of the angular momentum value of the excitation. The
combined effect of the Coulomb and spin-orbit interactions is
essentially given by the algebric sum of the two effects; the
interference phenomena are almost negligible.

The results obtained for the excitations dominated by the neutron
s.p. pairs indicated in Table \ref{tab:sp} are shown in panel (b) of
Fig. \ref{fig:pnmult}. In this case, the effect of the Coulomb
interaction is almost zero. It is not exactly zero since in the RPA
solution, even if dominated by the neutron transition, there are
contributions of some proton p-h pairs.  As in the proton case, we do
not identify general trends related to the inclusion of the spin-orbit
interaction. The size of the spin-orbit effects is not always
negligible and we observe differences of more than 
0.8 MeV for the $3^-$ of $^{60}$Ca and the $2^+$ of $^{78}$Ni.

It is worth noting (see panel (b) of Fig. \ref{fig:pnmult}) that those
nuclei with the same neutron number show a similar trend in the energy
differences for the neutronic excitations. This is apparent for
$^{78}$Ni and $^{90}$Zr and for the two common excitations of
$^{60}$Ca and $^{68}$Ni.

%%%%%%%%%%%%%%%%%%%%%%%%%%%%%

\section{Conclusions}   
\label{sec:conclusions}
In this paper we have presented the results of a study 
focused on the role of Coulomb and spin-orbit interactions in RPA
calculations. The inclusion of these two terms of the force 
is required when the s.p. wave functions and energies used
in RPA are generated by a HF calculation, to have a complete
self-consistency. 

We have conducted our investigation in various spherical
nuclei covering different regions of the nuclear chart  
by selecting some low-lying states where the details of 
the s.p. wave functions around the Fermi surface are relevant.

By studying the low-lying $2^+$ and $3^-$ excitations we found the
need of providing a proper treatment of the exchange term of the
Coulomb interaction, usually simulated by the Slater
approximation. The exchange term of the Coulomb interaction in RPA
calculations generates a globally attractive effect, i. e. an effect
which lowers the value of the energies calculated without it.  The use
of the direct term only has an opposite effect, and the Slater
approximation cannot solve the problem.

We have investigated the effects of the Coulomb interaction in IS and
IV excitations, and also in excitations dominated by specific p-h pairs
and coupled to different angular momentum values. 
The results of all our calculations confirm 
the attractive character of the Coulomb interaction.
The size of the effects of the Coulomb force is rather 
similar in all the cases we have investigated and it is of about few 
hundreds  of keV. 

The situation regarding the spin-orbit force is much more complicated. 
The study of the low-lying $2^+$ and $3^-$ states does not show
general behavior of the spin-orbit effects, even though in the great majority
of the cases the total effect is attractive. We found a generally larger 
sensitivity to the IS transitions than to the IV ones, however, we did 
not observe a general trend of the effects. The sign of the effect changes 
depending on the nucleus and on the multipolarity investigated. We observed 
an analogous situation also when we investigated 
excitations related to specific s.p. pairs. Also in this case, sign
and size of the  
effect depend on the nucleus investigated and on the angular momentum 
of the excitation. 

The effects of the spin-orbit interaction are larger than those of the Coulomb 
interaction. The energy difference with respect to the results obtained 
without them can be larger than 1 MeV. 
The size of the total effect obtained by including both Coulomb 
and spin-orbit forces in a unique RPA calculation 
is essentially given by the sum of the two separated results. 
Interference effects are negligible. 

We have also studied the possibility of identifying Coulomb and
spin-orbit effects in other observables, different from the excitation
energies, but we found very small effects, therefore we did not show
here these results.

Self-consistent mean-field models are the starting ground to make predictions
about the structure of exotic nuclei. The total self-consistency of these calculations
is a requirement which reinforces the reliability of these calculations. The exclusion
of the Coulomb and spin-orbit terms of the interaction can generate errors up to about 
1 MeV on the RPA excitation energies.

%%%%%%%%%%%%%%%%%%%%%%%%%%%%%%%%%%%%%
\appendix
\section{The spin-orbit matrix elements} 
\label{sec:appa}

In this appendix, we give some details of our calculations of the 
RPA matrix elements for the spin-orbit interaction terms $\widehat{V}_7$ and
$\widehat{V}_8$ in Eqs. (\ref{eq:force1}) and (\ref{eq:fcahnnels}). 
We calculate, first, the generic matrix elements
\beq 
\langle \widehat{V}_7\rangle \, \equiv \,
\langle (ab) J_1 M_1 | \widehat{V}_7(1,2) | (cd) J_2 M_2\rangle 
\, \langle (\frac{1}{2}t_a) \, (\frac{1}{2}t_b) | \mathbb{1} | (\frac{1}{2}t_c)Ê\, (\frac{1}{2}t_d) \rangle\, ,
\label{eq:me}
\eeq
where $a$, $b$, $c$ and $d$ indicate all the quantum numbers
identifying the s.p. states involved: the principal quantum number $n$, the
orbital angular momentum $l$ and the total angular momentum $j$.  
In addition, $t_i= \pm1/2$ indicates the third component of the
isospin of the $i$-th s.p. state. 
The explicit expression of the spin-orbit channel of the interaction, $\widehat{V}_7$, is:
\beq 
\widehat{V}_7(1,2) \, =\, v_7(r_{12})\, {\bf L}_{12}\cdot {\bf S} \, .
\label{eq:VLS}
\eeq 
From the definitions (\ref{eq:r})-(\ref{eq:p}),
 ${\bf L}_{12}$ can be expressed as:
\beq
{\bf L}_{12}=\frac{1}{2}
({\bf L}-\br_1 \times \bpi_2 -\br_2 \times \bpi_1) \, ,
\eeq
where 
\beq 
{\bf L}\, = \, {\bf l}_{1}+{\bf l}_{2} \, ,
\label{eq:Lt}
\eeq
with ${\bf l}_i= \br_i \times \bpi_i$. 
We express the matrix element of (\ref{eq:me}) as:
\beq
\langle \widehat{V}_7\rangle  \,
= \, \frac{1}{2}\, \left( V_{LS} \,- \, V_{12} \, -\, V_{21} \right) \, \delta_{t_a,t_c}\, \delta_{t_b,t_d} \, ,
\eeq
where
\beq
V_{LS} \, =\, \langle
(ab)J_1 M_1 |  v_7(r_{12})\, \mathbf{L}\cdot \mathbf{S} | (cd)J_2 M_2\rangle  \, ,
\label{eq:VLS2}
\eeq
\beq
V_{ij}\,=\, \langle
(ab)J_1 M_1 | v_7(r_{12})\, \mathbf{r}_i\times\mathbf{p}_j\cdot \mathbf{S} | (cd)J_2 M_2\rangle  
\label{eq:Vij}
\eeq
and the $\delta$'s come from the isospin matrix element.

As suggested in Ref. \cite{nak02}, we calculate $V_{LS}$ by changing from $jj$ to $LS$ coupling scheme:
\beqn
\nonumber
V_{LS} &=& 
\sum_{L_1S_1L_2S_2} \, \hat{j_a} \, \hat{j_b} \, \hat{L_1} \, \hat{S_1} \, \hat{j_c} \, \hat{j_d} \, \hat{L_2} \, \hat{S_2} \, 
\ninej {l_a}  \half  {j_a}  {l_b}  \half  {j_b}  {L_1}  {S_1}  {J}  \, 
\ninej {l_c}  \half  {j_c}  {l_d}  \half  {j_d}  {L_2}  {S_2}  {J} 
\\&~&
\langle(l_a l_b)L_1,(\half\half)S_1;J_1M_1| v_7(r_{12})\, \mathbf{L}\cdot \mathbf{S} | (l_cl_d)L_2,(\half\half)S_2;J_2M_2\rangle
\,\,,
\label{eq:accLS}
\eeqn
where we have used the Wigner 9-j symbol and, for the angular momentum
indexes, the convention $\hat{l} = \sqrt{2l+1}$. This coupling scheme
is convenient because the states are eigenstates of the
$\mathbf{L}\cdot\mathbf{S}$ operator. 
By using this property, we obtain:
\begin{eqnarray}
V_{LS}  & = & \sum_{L_1S_1L_2S_2} 
\, \hat{j_a} \, \hat{j_b} \, \hat{j_c} \, \hat{j_d} \, \hat{L_1} \, \hat{S_1} \, \hat{L_2} \, \hat{S_2} \, 
\frac{J_2(J_2+1)-L_2(L_2+1)-S_2(S_2+1)}{2} \, \ninej{l_a}{\half}{j_a}{l_b}{\half}{j_b}{L_1}{S_1}{J_1}
\nonumber\\
&&  
\ninej{l_c}{\half}{j_c}{l_d}{\half}{j_d}{L_2}{S_2}{J_2}  \,
\langle (l_al_b) L_1, (\half \half)S_1; J_1 M_1 | v_7(r_{12}) | (l_cl_d) L_2, (\half \half)S_2;J_2 M_2\rangle \, .
\end{eqnarray}

If $S_2=0$, $J_2=L_2$ and $J_2(J_2+1)-L_2(L_2+1)-S_2(S_2+1)=0$. Thus, the only term contributing to the sum on $S_2$ is 
$S_2=1$. Taking this into account we have:
\begin{eqnarray}
V_{LS}  & = & \sum_{L_1S_1L_2} 
\, \hat{j_a} \, \hat{j_b} \, \hat{j_c} \, \hat{j_d} \, \hat{L_1} \, \hat{S_1} \, \hat{L_2} \, \sqrt{3} \, 
\frac{J_2(J_2+1)-L_2(L_2+1)-2}{2} \, \ninej{l_a}{\half}{j_a}{l_b}{\half}{j_b}{L_1}{S_1}{J_1}
\nonumber\\
&&  
\ninej{l_c}{\half}{j_c}{l_d}{\half}{j_d}{L_2}{1}{J_2}  \,
\langle (l_al_b) L_1, (\half \half)S_1; J_1 M_1 | v_7(r_{12}) | (l_cl_d) L_2, (\half \half)1;J_2 M_2\rangle \, .
\end{eqnarray}
Now it is useful to consider the expansion in terms of spherical
harmonics $Y_{\lambda \mu}$: 
\beq
v_{7}(r_{12}) \, = \, 4\sqrt{2\pi} \sum_{L} (-1)^L \, \hat{L} \,  {\cal V}_L^{\rm (7)}(r_1,r_2) \, [Y_L(r_1) \otimes Y_{L}(r_2) ]^0_0\, ,
\label{eq:vls}
\eeq
where
\begin{equation}
 {\cal V}_L^{(7)}(r_1,r_2) \,=\, \int {\rm d}q \, q^2Ê\, j_L(qr_1) \, j_L(qr_2) \, \tilde{v}_{7}(q) \, . 
\end{equation}
The relation between $v_{7}(r_{12})$ and $\tilde{v}_{7}(q)$ is given
in eq. (\ref{force:ft}). Now we can decouple the $LS$
states and obtain: 
\begin{eqnarray}
V_{LS} 
&=& \frac{3}{\sqrt{2\pi}} \, \delta_{J_1,J_2} \, \delta_{M_1,M_2} \sum_{L_2L} (-1)^{L_2+L} \, \hat{l_a}\, \hat{l_b}\, \hat{l_c}\, \hat{l_d}\,\hat{j_a} \, \hat{j_b} \, \hat{j_c} \, \hat{j_d} \,\hat{L_2^2} \, \hat{L^2} \, [J_2(J_2+1)-L_2(L_2+1)-2] \nonumber \\
&& \ninej{l_a}{\half}{j_a}{l_b}{\half}{j_b}{L_2}{1}{J_2} \,
\ninej{l_c}{\half}{j_c}{l_d}{\half}{j_d}{L_2}{1}{J_2}  \, \sixj{l_a}{l_b}{L_2}{l_d}{l_c}{L} \, \threej{l_a}{l_c}{L}{0}{0}{0} \, \threej{l_b}{l_d}{L}{0}{0}{0} \nonumber \\
&& \int {\rm d}r_1 \, r_1^2 \, R_a^*(r_1) \, R_c(r_1) \int {\rm d}r_2 \, r_2^2 \, {\cal V}_L^{\rm (7)}(r_1,r_2) \, R_b^*(r_2)  \,
 \, R_d(r_2) \, .
\label{eq:v1-fin}
\end{eqnarray}

For the calculation of
$V_{12}$ we use again the $LS$ coupling scheme and the operator written
as follows: 
\beq
{\bf r_1}\times {\bf p}_2 \cdot {\bf S} \, = \, i\, \sqrt{6} \, [[{\bf r}_1 \otimes {\bf p}_2]^1 \otimes {\bf S}]^0_0 \, .
\eeq
The matrix element $V_{12}$  is given by:
\begin{eqnarray}
V_{12} & = &i\, \sqrt{6}   \sum_{L_1S_1L_2S_2} 
\, \hat{j_a} \, \hat{j_b} \, \hat{j_c} \, \hat{j_d} \, \hat{L_1} \, \hat{L_2} \, \hat{S_1} \, \hat{S_2} \, 
\ninej{l_a}{\half}{j_a}{l_b}{\half}{j_b}{L_1}{S_1}{J_1} \,
\ninej{l_c}{\half}{j_c}{l_d}{\half}{j_d}{L_2}{S_2}{J_2}   \\ \nonumber
&& \langle (l_al_b) L_1, (\half \half)S_1; J_1 M_1 | v_{7}(r_{12})\, [[{\bf r}_1 \otimes {\bf p}_2]^1 \otimes {\bf S}]^0_0 | (l_cl_d) L_2, (\half \half)S_2;J_2 M_2\rangle 
\end{eqnarray}

Decoupling the $LS$ states we obtain:
\begin{eqnarray}
V_{12} & = & \nonumber
 i\, \sqrt{2}  \, \delta_{J_1,J_2}\, \delta_{M_1,M_2}\,  \sum_{L_1S_1L_2S_2} (-1)^{L_2+S_1+1+J_2} \, 
\hat{j_a} \, \hat{j_b} \, \hat{j_c} \, \hat{j_d} \, \hat{L_1} \, \hat{L_2} \, \hat{S_1} \, \hat{S_2} \,  \\ \nonumber
&&
\ninej{l_a}{\half}{j_a}{l_b}{\half}{j_b}{L_1}{S_1}{J_2} \,
\ninej{l_c}{\half}{j_c}{l_d}{\half}{j_d}{L_2}{S_2}{J_2}\, \sixj{L_1}{S_1}{J_2}{S_2}{L_2}{1} 
\nonumber \\
&& 
 \langle (l_al_b) L_1 \| v_{7}(r_{12}) \, [{\bf r}_1 \otimes {\bf p}_2]^1 \| (l_cl_d) L_2\rangle \,
\langle (\half \half)S_1\| {\bf S}  \| (\half \half)S_2\rangle 
\, .
\label{eq:v2-2}
\end{eqnarray}
By using again the expansion of eq. (\ref{eq:vls}) we have:
\begin{eqnarray}
v_7(r_{12})\, [{\bf r}_1 \otimes {\bf p}_2]^1_\mu & = &
-i\,4\sqrt{2\pi} \sum_{LK_1K_2}   \hat{L} \,  \hat{K_1} \, \hat{K_2} \,   \sixj{L}{1}{K_1} {1}{K_2}{1}
\,  \threej{L}{1}{K_1} {0}{0}{0} 
\nonumber \\
&&
 r_1 \,  {\cal V}_L^{\rm (7)}(r_1,r_2) \,
[Y_{K_1}(r_1) \otimes [Y_{L}(r_2) \otimes \nabla_2]^{K_2}]^1_\mu
\, ,
\end{eqnarray}
where $\nabla$ is the gradient operator, 
and by considering that the spin matrix element is:
\beq
\langle (\half \half)S_1\| {\bf S}  \| (\half \half)S_2\rangle \, = \, \sqrt{6}\,  \delta_{S_1,S_2}\, \delta_{S_1,1}
\eeq
we have: 
\begin{eqnarray}
V_{12} 
&=& 18\sqrt{\frac{2}{\pi}}  \, \delta_{J_1,J_2}\, \delta_{M_1,M_2}\,  \sum_{L_1L_2} \sum_{LK_1K_2}  (-1)^{l_a+l_d+L_2+J_2} \, \hat{l_a} \, \hat{l_b} \,  \hat{l_c} \, \hat{l_d} \, 
\hat{j_a} \, \hat{j_b} \, \hat{j_c} \, \hat{j_d} \, \hat{L_1^2} \, \hat{L_2^2} \, \hat{L^2} \,  \hat{K_1^2} \, \hat{K_2^2}  
\nonumber \\
&& 
\ninej{l_a}{l_b}{L_1}{l_c}{l_d}{L_2}{K_1}{K_2}{1} \, \ninej{l_a}{\half}{j_a}{l_b}{\half}{j_b}{L_1}{1}{J_2} \,
\ninej{l_c}{\half}{j_c}{l_d}{\half}{j_d}{L_2}{1}{J_2} \,\sixj{L_1}{1}{J_2}{1}{L_2}{1} 
\,  \sixj{L}{1}{K_1} {1}{K_2}{1}
 \nonumber \\
 &&
 \threej{L}{1}{K_1} {0}{0}{0} \, \threej{l_a}{K_1}{l_c}{0}{0}{0} 
 \int {\rm d}r_1 \, r_1^3 \, R_a^*(r_1) \, R_c(r_1) \,
  \int  {\rm d}r_2 \, r_2^2 \, R_b^*(r_2) \,  {\cal V}_L^{\rm (7)}(r_1,r_2)
     \nonumber \\
 &&
\left\{ \threej{l_d}{K_2}{l_b}{0}{0}{0} \, \threej{K_2}{1}{L}{0}{0}{0} \,
 \frac{\rm d}{{\rm d}r_2} \, R_d(r_2) \right. \\ \nonumber  
&& \left. +\,  \sqrt{2} \, \xi(l_b+l_d+L+1) \, \sqrt{l_d(l_d+1)}Ê\, \threej{l_d}{K_2}{l_b}{1}{-1}{0} \, \threej{K_2}{1}{L}{1}{-1}{0} \,  
 \frac{1}{r_2} \, R_d(r_2)  \right\}
 \, ,
\label{eq:v2-4}
\end{eqnarray}
where $\xi(n)=1$ or 0 if $n$ is even or odd, respectively.

The expression of $V_{21}$ can be obtained by using the same
procedure with the obvious changes. 

For the isospin dependent channel,
\beq
\widehat{V}_8(1,2) \, \equiv \, v_8(r_{12})\, {\bf L}_{12}\cdot {\bf S} \, \btau(1)\cdot\btau(2) \, ,
\label{eq:VLSt}
\eeq
the calculation is similar with
the only difference of the isospin matrix element which is now:
\begin{eqnarray}
\langle (\frac{1}{2}t_a) \, (\frac{1}{2}t_b) | \btau(1)\cdot\btau(2) | (\frac{1}{2}t_c)Ê\, (\frac{1}{2}t_d) \rangle & = & \\
&& \hspace*{-3cm} 2 \, \left( \delta_{t_a,\half}\, \delta_{t_b,-\half} \,\delta_{t_c,-\half}\, \delta_{t_d,\half} \, + \, \delta_{t_a,-\half}\, \delta_{t_b,\half} \,\delta_{t_c,\half}\, \delta_{t_d,-\half} \right) \, + \nonumber \\
&& \hspace*{-2.6cm}  \left(\delta_{t_a,\half}\, \delta_{t_c,\half}\,-\,\delta_{t_a,-\half} \, \delta_{t_c,-\half}\right) 
\left(\delta_{t_b,\half}\, \delta_{t_d,\half}\,-\,\delta_{t_b,-\half} \, \delta_{t_d,-\half}\right)  \nonumber
\, .
\end{eqnarray}

%%%%%%%%%%%%%%%%%%%%%%%%%%%%%%%%%%%%%%%%%%%%%%%%%%%%%%%%%%%%%%%%%%%%%%%%
\acknowledgments 
This work has been partially supported by the PRIN
(Italy) {\sl Struttura e dinamica dei nuclei fuori dalla valle di stabilit\`a}, by the Junta de
Andaluc\'{\i}a (FQM0220) and European Regional Development Fund (ERDF) and the Spanish Ministerio de Econom\'{\i}a y Competitividad (FPA2012-31993).

%%%%%%%%%%%%%%%%%%%%%%%%%%%%%%%%%%%%%%%%%%%%%%%%%%%%%%%%%%%%%%%%%%%%%%%%%
%   Bibliography
%%%%%%%%%%%%%%%%%%%%%%%%%%%%%%%%%%%%%%%%%%%%%%%%%%%%%%%%%%%%%%%%%%%%%%%%%
%\newpage
%\bibliographystyle{elsart-num} 
%\bibliography {altro,bcs,eweak,exp,gpco,hyp,libri,micro,rpa,tesi,nuovo}

\end{document}